astro-ph/9512129 19 Dec 1995

# The Evolution of Viscous Inclined Disks in Axisymmetric and Triaxial Galaxies

Wesley N. Colley [1]
Princeton University Observatory, Peyton Hall, Princeton, NJ 08544

Linda S. Sparke
University of Wisconsin, Washburn Observatory, 475 N. Charter St., Madison, WI 53706

## ABSTRACT

We have used a set of equations developed by Pringle (1992) to follow the evolution of a viscous twisted disk in a galaxy-like potential which is stationary or tumbling relative to inertial space. In an axisymmetric potential, the disk settles to the equatorial plane at a rate determined largely by the coefficient $\nu_2$ associated with shear perpendicular to the local disk plane. If the disk is initially close to the galaxy equator, the rate at which the inclination decays is well described by the analytic formula of Steiman-Cameron & Durisen (1988); in a highly inclined disk, 'breaking waves' of curvature steepen as they propagate through the disk, rendering the numerical treatment untrustworthy. In a triaxial potential which is stationary in inertial space, settling is faster than in an oblate or prolate galaxy, since the disk twists simultaneously about two perpendicular axes. If the figure of the potential tumbles about one of its principal axes, the viscous disk can settle into a warped state in which gas at each radius follows a stable tilted orbit which precesses so as to remain stationary relative to the underlying galaxy.

## 1. Introduction

Warped and twisted disks are common in the galactic context. In most spiral galaxies, the neutral hydrogen disk is warped out of the plane of the central galaxy, often by tens of degrees. Models for galactic warps have been primarily gravitational in nature, ignoring gas processes within the disk; see Binney (1992) for a review. Early-type galaxies usually contain little gas, but when gas is present, it is often found in fairly ordered disks or rings, orbiting at a large angle to the apparent symmetry planes of the underlying galaxy; the polar rings catalogued by Whitmore *et al.* (1990) are striking examples. On a smaller scale, the centers of galaxies often contain warped or tilted gas disks: within the central kiloparsec of our own Milky Way, HI and molecular gas follow orbits tipped with respect to the outer disk (*e.g.*, Burton 1992); HII regions in the central bulge of M31 occupy a tilted disk (Ciardullo *et al.* 1988); and a number of barred galaxies including NGC 2217 (Bettoni, Fasano & Galletta 1990) have warped and twisted central disks of gas and dust.

[1] Supported by the Fannie and John Hertz Foundation, Livermore, CA 94551-5032

These tilted gas disks generally appear thin and have a regular velocity structure, which implies that they have settled dissipatively at least part-way towards an equilibrium state. Their evolution presents a problem in three-dimensional hydrodynamics, so that any theoretical investigation necessarily involves many approximations. Christodoulou & Tohline (1991) adopted a hydrodynamic approach to gas disks in an oblate potential; even with a fairly coarse computational grid, their simulations required the use of a supercomputer. Particle-based techniques, such as the smoothed-particle scheme developed by Lucy (1977), allow high resolution where material is present, while avoiding the waste of machine time in following empty regions of the computational volume. Such schemes have been used by Tubbs (1980), Hernquist & Katz (1989), Varnas (1990) and Katz & Rix (1992) to follow the time-development of gas disks in oblate and prolate potentials, while Habe & Ikeuchi (1985) used them to study the settling of disks in prolate and triaxial potentials. In these schemes, the time-step is limited by the need to follow the orbital motion of individual fluid particles, while the inflow and settling perpendicular to the local disk plane are much slower. Steiman-Cameron & Durisen (1988) developed a system of analytic equations describing the time-development of a twisted viscous gas disk, assuming the disk to be in centrifugal balance and thus removing the need to integrate the fast orbital motion, which they used to investigate the timescale on which an initially tilted gas disk would settle to the equatorial plane of an oblate galaxy. Such an orbit-averaging method has the limitation that it can follow only motions which are slow compared to the orbital velocities, and could not treat the collapse and rapid infall of a gas disk.

Investigations of the time-development of twisted galactic disks are also hampered by our ignorance of the nature of the dissipation. Molecular viscosity is negligible, and the complex dissipative processes in the interstellar medium cannot be simply characterized. In modelling the disk dynamics, one generally develops a mathematically simple description of viscosity as a process which dissipates energy and causes momentum to diffuse, and hopes that the results are at least qualitatively independent of the details of that process. In the context of planetary rings, we know that this hope is false – some forms of dissipation make the rings spread, and others cause them to become narrower (see, *e.g.*, Section 5.3.2 of Goldreich & Tremaine 1982, Wisdom & Tremaine 1988). Comparing the predictions of the various numerical schemes provides a test which of the effects found in simulations are dependent on the forms of viscosity used; behavior found in common may be generic, and indicative of what is to be expected in real galaxies. Christodoulou *et al.* (1992) make such a comparison of hydrodynamic and particle-based simulations, and note various points of agreement.

Pringle (1992) has recently developed an orbit-averaging scheme in which gas inflow and twisting are controlled by two viscous coefficients; the first governing forces depending on shear in the plane of the disk, while the second controls forces perpendicular to the local disk. Pringle's formulation is somewhat simpler than that of Steiman-Cameron & Durisen (1988), and has the additional advantage that it explicitly conserves local angular momentum. Here, we apply Pringle's equations in discretized form to investigate the time-evolution of warped and tilted gas disks in axisymmetric and triaxial galactic potentials. We calculate rates of settling into equilibrium states, examine the inflow associated with settling, search for long-lived warped states, and compare our results with those of hydrodynamic and particle-based simulations. The equations are briefly described in Section 2 below. Section 3 then considers the settling of a gas disk initially tilted away from the symmetry plane of an axisymmetric potential; in Section 4 we discuss triaxial potentials held stationary in inertial space, and in Section 5, tumbling triaxial potentials; Section 6 summarizes our results and conclusions.

## 2. Equations of Motion

Following Pringle (1992), we consider a thin twisted disk such that material is centrifugally supported in circular orbits, with the pole at distance $R$ from the center pointing along the unit normal $\mathbf{l}(R,t)$; the disk is taken to be thin in the perpendicular direction. We assume that the orbital angular velocity $\Omega(R)$ around the orbit is set by the gravitational potential of an underlying mass distribution, and that the disk contributes negligibly to the gravitational forces. The annulus of material between radii $R - \Delta R/2$ and $R + \Delta R/2$ then has angular momentum $\Delta\mathbf{L} = 2\pi R\mathbf{L}DeltaR \approx 2\pi R\Sigma \cdot \Delta R \cdot R^2\Omega(R)\mathbf{l}$, where $\Sigma(R,t)$ is the surface mass density of the disk. The evolution of the angular momentum per unit area $\mathbf{L}$, under the influence of viscosity and of an external torque $\mathbf{N}(R,\mathbf{L},t)$, is then given by Pringle's equation (2.6):

$$\frac{\partial}{\partial t}\mathbf{L}(R,t) = \frac{1}{R}\frac{\partial}{\partial R}\left[\frac{(\partial/\partial R)[\nu_1 \Sigma R^3(-\Omega')]}{\Sigma(\partial/\partial R)(R^2\Omega)}\mathbf{L}\right] + \frac{1}{R}\frac{\partial}{\partial R}\left[\frac{1}{2}\nu_2 R|\mathbf{L}|\frac{\partial \mathbf{l}}{\partial R}\right] + \frac{1}{R}\frac{\partial}{\partial R}\left\{\left[\frac{1}{2}\frac{\nu_2 R^3\Omega|\partial\mathbf{l}/\partial R|^2}{(\partial/\partial R)(R^2\Omega)} + \frac{\nu_1 R\Omega'}{\Omega}\right]\mathbf{L}\right\} + \mathbf{N}(R,\mathbf{L},t)\ , \tag{1}$$

where $\Omega' \equiv d\Omega/dR$. The two coefficients $\nu_1$ and $\nu_2$ correspond to viscosity acting on the azimuthal and vertical shear respectively; thus $\nu_1$ is the shear viscosity normally associated with planar accretion disks, while $\nu_2$ controls the damping of out-of-plane motions associated with a twist in the disk. Since $\Omega(R)$ is given, the surface density $\Sigma(R,t)$ is then specified by the magnitude of $\mathbf{L}$.

This first-order equation clearly conserves angular momentum when external torques are absent. The derivation assumes that the angular momentum of each ring lies entirely along the normal to the orbital plane, so equation (1) describes only the slow precession of the disk, and does not follow fast nutational motions; the rate of change of angular momentum due to the external torque $\mathbf{N}$ should be small enough not to violate this assumption. ??? For a flat disk, with $\partial\mathbf{l}/\partial R = 0$, the equation reduces to the standard form for a flat disk (*e.g.*, Pringle 1981). The first two terms are diffusive, while the third is advective, with the advective velocity $V_{adv}$ given by

$$V_{adv} = -\frac{1}{2}\frac{\nu_2 R^3\Omega|\partial\mathbf{l}/\partial R|^2}{R(\partial/\partial R)(R^2\Omega)} - \frac{\nu_1\Omega'}{\Omega}\ . \tag{2}$$

If the disk is flat, or if $\nu_2 = 0$, $V_{adv}$ is positive; it becomes negative when the curvature $d\mathbf{l}/dR$ is strong.

To discretize equation (1), we follow Pringle (1992) in using a first-order explicit scheme which treats the Cartesian $(x,y,z)$ components of the angular momentum separately, spacing our grid points evenly in radius. The standard Forward Time Centered Space differencing is used for the two diffusive terms with zero-torque boundary conditions at the inner and outer edges of the radial grid, as outlined by Pringle, so that angular momentum is conserved exactly under the action of the first two terms of equation (1). The third, advective, term is treated by upstream differencing (*e.g.*, Press *et al.* 1992), so that angular momentum is lost or gained if $V_{adv}$ changes sign in the middle of the grid; this can be a problem in the settling disks that we study here. The zero-torque boundary conditions require $d\mathbf{l}/dR = 0$ at the inner and outer edges of the disk, so the advective velocity is always positive there; angular momentum (and hence mass) flows outwards off the grid at a rate proportional to the local density. Therefore we choose the outer boundary to lie well beyond most of the mass of the disk. The timestep $\Delta t$ must be chosen short enough to satisfy the Courant condition on the diffusive terms:

$$\Delta t < (\Delta R)^2 \left[ \max \left( \frac{\nu_1 R^2(-\Omega')}{(\partial/\partial R)(R^2\Omega)}, \frac{1}{2}\nu_2 \right) \right]^{-1} \quad . \tag{3}$$

Thus the integration scheme is mathematically simple but computationally is relatively inefficient.

It is not at all clear what the appropriate viscosity for a galactic gas disk should be. An estimate may be made by considering a disk made up of gas clouds which collide with mean free path $\lambda$; the coefficients of kinematic viscosity are then given by $\nu \sim \lambda v_1$, where $v_1$ is the one-dimensional cloud velocity dispersion (*e.g.*, Shu 1992). The disk is thin, so that if the cloud velocity dispersion is close to isotropic, the clouds have random speeds well below the local circular velocity: $v_1 \ll R\Omega(R)$. The maximum distance they can move between collisions is then given by the epicycle amplitude: $\lambda < v_1/\kappa$ where $\kappa$ is the epicyclic frequency (Binney & Tremaine 1987); when the rotation curve is flat, with constant linear speed $V$, $\kappa = \sqrt{2}V/R$. This sets an upper limit to the viscosity: $\nu \leq \nu_{max} \sim v^2{}_{rms}R/\sqrt{6}V$, where $v_{rms}$ is now the three-dimensional cloud velocity dispersion; with Steiman-Cameron & Durisen (1988), we adopt

$$\nu_{max} = 0.2 v^2{}_{rms} R/V \ . \tag{4}$$

We choose units such as to measure lengths in kiloparsecs and velocities in units of 100 km s$^{-1}$; our time unit is thus about $10^7$ years. Typical velocity dispersions in the cold H I layer of our Galaxy and others are about $5 - 10$ km s$^{-1}$ (Burton 1992); taking $V_{rms} = 10$ km s$^{-1}$, $R = 5$ kpc and a rotation speed of 250 km s$^{-1}$, the viscosities $\nu_1$ and $\nu_2$ should not exceed about $4 \times 10^{-3}$.

## 3. Oblate or Prolate Galaxy Potential

In an axisymmetric galaxy potential, the precessional torque may be written as $\mathbf{N} = \Omega_p(R, \mathbf{l})\mathbf{z} \times \mathbf{L}$, where the unit vector $\mathbf{z}$ points along the polar axis of the galaxy. The scale-free models of Richstone (1980; see also Toomre 1982) describe axisymmetric potentials in which the linear speed $V$ of orbital motion is independent of radius: $\Omega(R) = V/R$. Richstone's model, which Toomre labels by $(n = 0, m = 1)$, has a potential $\Phi$ which is constant on concentric similar spheroids: at (spherical) radius $R$ and height $z$ above the equatorial plane,

$$\Phi(R, z) = \frac{V^2}{2} \ln \left\{ \frac{R^2}{R_0{}^2} [1 + \eta P_2(z/R)] \right\} \approx V^2 \left\{ \ln \frac{R}{R_0} + \frac{\eta}{2} P_2(z/R) \right\} \tag{5}$$

where $P_2$ is the Legendre polynomial, and the parameter $\eta$ controls the asphericity; positive $\eta$ corresponds to an oblate mass distribution and $\eta < 0$ to a prolate system, and the approximate equality holds when $\eta$ is small. The scaling radius $R_0$ does not appear in the expressions for forces and torques. Averaging over a uniform circular ring at radius $R$ with inclination $i = \arccos(\mathbf{l} \cdot \mathbf{z})$ to the equatorial plane yields a torque per unit mass

$$\mathbf{N} = -\frac{3V}{4R}\eta(\mathbf{z} \cdot \mathbf{l})(\mathbf{z} \times \mathbf{L}) \ . \tag{6}$$

If viscosity is absent, this will cause the orbits of material at radius $R$ to precess about the $z$-axis at the rate

$$\Omega_p(R, i) = -\frac{3V}{4R}\eta(\mathbf{z} \cdot \mathbf{l}) \tag{7}$$

(*cf.*, Sparke 1986); if $\cos i \propto R$ then the entire warped disk precesses rigidly without change of shape.

The Smoothed Particle Hydrodynamic simulations of Katz, as reported in Section 2.2 and Figure 1 of Christodoulou *et al.* (1992), followed a near-polar ring initially inclined by 80° to the equatorial plane of both an oblate and a prolate potential. In the oblate potential, the ring locked into a steadily precessing state, warping down towards the equatorial plane with inclination approximately following the relation $\cos i \propto R$. This warped state persisted for at least 30 orbital periods. In the prolate potential, the ring behaved very differently; it initially developed a warp up towards the pole, after which the material collapsed within five rotation periods into the center. The addition of self-gravity to the ring material did not change these results. Interestingly, the sense of warping is exactly opposit to that predicted when considering self-gravity alone; the stable configuration for a massive near-polar ring in an oblate galaxy is one warping up towards the polae at larger radii (Sparke 1986, Arnaboldi & Sparke 1994), while in a prolate potential the stable state warps down towards the equator (Sparke 1995).

We can compare the results of Katz's particle-based simulations with the predictions of equation (1). In an axisymmetric galaxy, the precession-inducing torque has no component along the axis $z$. Under the interchange $(\mathbf{N}, x, y) \rightarrow (-\mathbf{N}, -x, -y)$, the $z$-component of equation (1) for $\partial \mathbf{L}/\partial t$ is unchanged, while the $x$ and $y$ components reverse their signs. Thus the inclination $i$ must evolve in a prolate potential exactly as a disk with twist angles mirror-reflected through the $z$-axis would evolve in the corresponding oblate potential with equal and opposite flattening. In particular, if the disk evolves to a steadily precessing state in an oblate potential, it should reach a corresponding steady state in a prolate potential, with the same inclination to the $z$-axis, but with reversed sense of precession and azimuthal twist. The effective viscous interaction in the computations described by Christodoulou *et al.* (1992) is a complex mixture of 'pressure' forces and cooling. The fact that those calculations show a difference between the ring evolution in oblate and prolate galaxy potentials, while equation (1) insists that the behavior must be the same but for a reversal of helicity, implies that this aspect of the disk evolution may be sensitive to the particular form of viscosity chosen.

Figure 1 shows, for an annulus of material settling from an initial inclination of 30° in a potential with $\eta = 0.2$, the inclination, the surface density, the speed of radial inflow $V_R$ calculated according to equation (2.3) of Pringle (1992), and the advective speed $V_{adv}$. The density $\Sigma(R)$ was initially given by a Gaussian, truncated at the inner and outer edges, which implies that the radial speed $V_R$ is not smooth at the inner and outer edges. We also tried runs where $\Sigma(R)$ is modified so that $V_R$ approaches zero smoothly at the edges, which requires the derivative of $\Sigma(R)R^3 d\Omega/dR$ to go to zero. Our results were almost identical, and since the condition required for smooth $V_R$ does not seem physically natural, we have used the truncated Gaussian form for $\Sigma(R)$.

As seen in Figure 1, the innermost rings settle fastest towards the equatorial plane. The non-monotonic behaviour at the inner edge is a consequence of the stress-free boundary condition: the ring at the edge feels a torque only from the penultimate ring, and so takes longer to settle than rings which experience torque from material on both sides. Inflow is slow: both the radial and advective speeds are less than 1% of the circular speed in the disk. For $t > 0$, the minima of $V_R$ reflect those of $V_{adv}$, which occur where the disk is actively settling and $d\mathbf{l}/dR$ is appreciable; inflow is most rapid in these regions. The advective speed is positive in the inner, settled part of the disk but becomes negative further out, so we lose about 2% of the mass and 1% of the $z$-angular momentum during the run.

Figure 2 shows the deviation of the azimuth of the line of nodes from that expected for a freely precessing disk with an inclination of 30°. At any given time, the orbits of material near the inner edge of the disk have regressed somewhat less than they would have in the absence of viscosity, because matter has flowed in from larger radii where regression is slower; in the outermost region, where material has moved outwards, more regression has taken place than if viscosity had been absent. But the deviations from free precession are small, consistent with the conclusions reached Section IIIb of Steiman-Cameron & Durisen (1988): as long as the viscosity is well below a critical level, here corresponding to $\nu_{crit} \simeq \eta V R$, viscous torques have little effect on the rate of precession, while viscosity strong enough to prevent differential precession will cause rapid inflow on the timescale characteristic of that precession.

Figure 3 shows the effect of reducing the in-plane viscosity $\nu_1$ by a factor of ten; settling proceeds much as before but inflow is much reduced. The sharp dip in both radial and advective speeds reflects a 'breaking wave' instability which we discuss further below; here it occurs in a region of the disk which contains very little mass and does not affect settling over the rest of the disk. When the 'twisting' viscosity $\nu_2$ is reduced, Figure 4 shows that the inclination decays more slowly, but inflow remains much as in Figure 1; at this low inclination, the inflow and settling are not strongly coupled.

Habe & Ikeuchi (1985) used a smoothed-particle hydrodynamic code to investigate the settling of an initially inclined annulus of gas in a prolate potential. They found that an annulus initially tilted by 10° settled to a vertically thin ring in the equatorial plane within ten rotation periods with little inflow, but a disk initially inclined at 80° collapsed to a third of its initial radial extent as it settled. Varnas (1990), also using a smoothed-particle code, found that inflow was more rapid in disks with a higher initial tilt, as did Steiman-Cameron & Durisen (1988), using their orbit-averaging equations. Figure 5a, for an initial inclination of 60°, indeed shows stronger inflow, with mass building up at the inner edge of the grid. The curves of radial and advective velocity are very similar to each other; radial motion is largely controlled by the local disk curvature $d\mathbf{l}/dR$, and strong inflow is confined to regions of the disk where the inclination changes rapidly.

However, we cannot give a reliable estimate of the inflow rate. Because $V_{adv}$ changes abruptly from positive to negative in the middle of the grid, 26% of the initial mass and 5% of the initial $z$-angular momentum in our calculation is lost. A finer computational grid causes the jump in $V_{adv}$ to steepen, as shown in Figure 5b, worsens the angular momentum loss, and hastens the inflow. We can see from equation (2) what has gone wrong: in regions of strong curvature, angular momentum is advected inwards at a faster rate, leading to a still greater steepening of the curvature, and a behaviour similar to the breaking of water waves (*e.g.*, Whitham 1974). The disk curvature grows to the limit set by the finite resolution of the radial grid, so that with a finer mesh the waves steepen still further and angular momentum conservation worsens. Reducing $\nu_2$ relative to $\nu_1$ does not help; only the $\nu_2$ viscosity can smooth out the twisting caused by differential precession, so that when it is small the disk builds up a large curvature and the same 'breaking waves' are seen. It appears that this set of equations cannot be used to treat the settling of disks which are initially at a large angle to the plane about which they precess. We found good results (mass conservation to a few percent and improving with a finer grid; flow quantities converging to limiting values with a finer grid) only for inclinations below about 40°.

We can compare the settling of our model disks with a simple analytic formula derived by Steiman-Cameron & Durisen (1990) in the limit that viscosity is weak. They showed that an initially tilted disk should settle towards the equatorial plane of an oblate galaxy potential approximately according to their equation (2). At any radius $R$, the inclination $i$ to the equatorial plane decays with a time constant $\tau_e$:

$$i \propto \exp[-(t/\tau_e)^3] \ , \quad {\tau_e}^{-3} = (\nu/6)(d\Omega_p/dR)^2 \ . \tag{8}$$

We recover their formula if we set $\nu = \nu_2$, $\nu_1 = 0$ in equation (1), and assume that the curvature in the ring is small, so that the diffusive term in $\nu_2$ is much larger than the advective term, which is quadratic in $|d\mathbf{l}/dR|$; we must further assume that subsequently the azimuth of the line of nodes of the ring changes much more rapidly with radius than any other quantities, and that the inclination is small so that $\Omega_p$ can be evaluated in the equatorial plane. In that case,

$$\frac{\partial}{\partial t}\mathbf{L}(R,t) \approx \frac{\nu_2}{2}\frac{\partial^2 \mathbf{L}}{\partial R^2} + \Omega_p(R, i=0)\mathbf{z} \times \mathbf{L}(R,t) \ . \tag{9}$$

In a freely-precessing disk which is initially coplanar, the radial derivative in equation (9) eventually grows as $-(d\Omega_p/dR)^2 t^2 \mathbf{L}$; writing the off-axis angular momentum as the product of the freely-precessing motion and a slow decay in the amplitude of the tilt, and equating components of equation (9) parallel to the off-axis component of $\mathbf{L}$ shows that $L \sin i$ decays according to equation (8).

Figure 6 compares the prediction of equation (8) to the computed decay in inclination for the disk of Figure 1, and for a similar run with both coefficients of viscosity reduced by a factor of ten. That prediction approximates the disk inclination well for low viscosity, but the fit is less good at the higher viscosity, when Figure 1 shows that appreciable inflow takes place.

Finally, Figure 7 shows the development of a disk from an initial state with a straight line of nodes, and inclination given by $\cos i \propto R$; in the absence of viscosity, this disk would precess rigidly without change of shape. Katz & Rix (1992), using smoothed-particle hydrodynamics to investigate highly inclined gas disks in an oblate galaxy potential, found that unless the random speeds of the particles were subjected to a strong 'cooling', particle collisions were frequent, and most of the material sank towards the center. When cooling was included, the gas settled into an annulus warped approximately according to $\cos i \propto R$, with only a small dispersion in velocity. Since particles in the cooled disk follow near-circular paths, the orbit-averaged equations discussed here ought to give a good representation of this behavior. But here we find that viscous diffusion is destabilizing; as material diffuses inwards and outwards, the inclination of the inner disk falls and that of the outer disk rises, and eventually sufficient differential precession occurs that the disk starts to settle and inflow becomes rapid. A larger viscosity increases the rate of both inflow and settling. According to the scheme considered here, an initially tilted disk can never settle into a state where $\cos i \propto R$; this aspect of the disk behavior may depend on the character of the dissipative scheme used in the computation.

## 4. Triaxial Galaxy Potential

To model a triaxial galaxy potential, we include a second torque term of the same form as in equation (6):

$$\mathbf{N} = -\frac{3V}{4R}[\eta_x(\mathbf{x}\cdot\mathbf{l})(\mathbf{x}\times\mathbf{L}) + \eta_z(\mathbf{z}\cdot\mathbf{l})(\mathbf{z}\times\mathbf{L})] \equiv \mathbf{\Omega_p}(R,\mathbf{L})\times\mathbf{L}\ . \tag{10}$$

Choosing $\mathbf{x}$ to lie along the longest axis of the potential, and $\mathbf{z}$ to lie along the shortest, we have $\eta_x < 0$ and $\eta_z > 0$. In the absence of viscosity, each ring now precesses about either the $x$ or the $z$ axis, depending on its initial orientation. Because the $x$ and $z$ components of $\mathbf{\Omega_p}$ have the same dependence on the respective components of $\mathbf{l}$, in free precession

$$\partial(\mathbf{\Omega_p}\cdot\mathbf{l})/\partial t = 0\ ; \tag{11}$$

the conserved quantity $\mathbf{\Omega_p}\cdot\mathbf{l}$ can be used to derive the precession trajectories of $\mathbf{l}$, shown in Figure 8. The extremal values of $\mathbf{\Omega_p}\cdot\mathbf{l}$ are $(\mp 3V\eta_z/4R)$, when the angular momentum is aligned or antialigned with the $z$-axis, and $(\mp 3V\eta_x/4R)$, when it lies along the $x$-direction. Closed loops of constant $\mathbf{\Omega_p}\cdot\mathbf{l}$ corresponding to precession about the $z$-axis circle the extremal point at $\mathbf{l} = (0,0,\pm 1)$; loops around $\mathbf{l} = (\pm 1,0,0)$ represent precession about the $x$-axis. The points $(0,\pm 1,0)$ are saddle points: there is no starting inclination for which the ring precesses about the intermediate $y$ axis, since particle orbits circling the intermediate axis of a stationary triaxial potential are unstable (Binney 1978, 1981; Heiligman & Schwarzschild 1979). Because $\mathbf{\Omega_p}$ is scale-free, the normals $\mathbf{l}$ of the rings making up an initially coplanar disk all follow the same precession trajectory, at rates which decrease inversely with radius.

Figure 9 shows the development of an initially coplanar disk with $\nu_1 = \nu_2 = 10^{-3}$ in a potential specified by $\eta_x = -0.2$, $\eta_z = 0.2$. The initial inclination is 30°, and the azimuth $\omega$ of the line of nodes of the disk, at which it cuts the $x$-$y$ plane, is 45°: $\mathbf{\Omega_p}\cdot\mathbf{l}$ is equal to 0.625 of its extremal value, and the disk precesses about the $z$-axis. The wavy pattern in the inclination reflects the fact that the free precession trajectories of Figure 8 do not maintain a fixed inclination as the azimuth changes; since the precession rate drops with radius, the azimuthal dependence appears as a radial variation. Observations of tilted gas disks which appear to have been twisted by precession [S0 galaxy NGC 4753 (Steiman-Cameron, Kormendy & Durisen 1992), NGC 660 (van Driel *et al.* 1995) and NGC 3718 (Schwarz 1985; see also Cox & Sparke 1996)] would allow one to limit the triaxiality of the potential, since the disk inclination should then depend systematically on azimuth. Steiman-Cameron, Kormendy & Durisen (1992) were able to model the warped dust lane in NGC 4753 as a twisted disk in which the inner gas had precessed by almost two complete revolutions relative to the outer material in the oblate potential of the galaxy, with no appreciable settling or noticeable dependence on azimuth, which suggests that the mass distribution in that galaxy is almost axisymmetric.

The rings making up the tilted disk of Figure 9 settle to the $x$-$y$ plane at a rate somewhat faster than in the oblate potential, because the disk is now being twisted about both $x$ and $z$ axes. The mean inclination is fairly well fit by the Steiman-Cameron & Durisen (1990) formula (8), with mean squared precession rate $(d\Omega_p/dR)^2$ set equal to twice the square of the mean $z$-precession derivative $(3V/4R^2)\eta_z \cos i$; the factor of two allows for twisting due to the $x$-torque. Inflow speeds are larger than in Figure 1; again, most of the inflow takes place during the time that the region of the disk which contains most of the mass is actively settling.

Figure 10 shows the development of a disk starting at $i = 80°$, $\omega = 45°$, which precesses about the $x$-axis and eventually settles into the $y$-$z$ plane. This initial state is somewhat further from the plane into which the disk settles, with $\mathbf{\Omega_p} \cdot \mathbf{l}$ equal to only 0.455 of the extremal value reached by a disk in the $y$-$z$ plane. The inflow is correspondingly faster; a hint of the 'breaking wave' instability is seen in the early stages of the calculation, and the conservation of mass is somewhat worse than for Figure 9. The formula (8) again gives a good estimate of the rate of settling.

## 5. Tumbling Triaxial Galaxy Potential

If the galaxy potential of Section 4 is taken to tumble about the axis $z$ at some rate $\Omega_t$, then the torque will contain a component normal to the instantaneous direction of the long axis of the potential, $\tilde{\mathbf{x}} = \mathbf{x} \cos \Omega_t t + \mathbf{y} \sin \Omega_t t$. We now have

$$\mathbf{N} = \mathbf{\Omega_p}(r, \mathbf{l}, t) \times \mathbf{L} = -\frac{3V}{4R}[\eta_x(\tilde{\mathbf{x}}.\mathbf{l})(\tilde{\mathbf{x}} \times \mathbf{L}) + \eta_z(\mathbf{z} \cdot \mathbf{l})(\mathbf{z} \times \mathbf{L})] \ , \qquad (12)$$

with $\eta_x < 0$, $\eta_z > 0$. In free precession with no viscous forces,

$$\frac{\partial(\mathbf{\Omega_p} \cdot \mathbf{l})}{\partial t} = -\frac{3V}{2R}\eta_x(\tilde{\mathbf{x}} \cdot \mathbf{l})\mathbf{l} \cdot \frac{\partial \tilde{\mathbf{x}}}{\partial t} = 2\Omega_t \mathbf{z} \cdot \frac{\partial \mathbf{l}}{\partial t} \ ; \qquad (13)$$

extremal values of $(\mathbf{\Omega_p} - 2\Omega_t \mathbf{z}) \cdot \mathbf{l}$ correspond to orientations $\mathbf{l}$ at which the ring precesses along with the figure of the potential.

The positions of these extrema depend on the tumble rate $\Omega_t$, which without loss of generality may be taken as positive. When tumbling is slow, $0 < \Omega_t < 4R/3V\eta_z$, the poles $\mathbf{l} = (0, 0, \pm 1)$ are both stable fixed points of the precessional motion, while the saddle points, which were at $\mathbf{l} = (0, \pm 1, 0)$ in the static potential, now migrate towards the retrograde pole, and are at $l_x = 0, l_z = -4R\Omega_t/3V\eta_z$. The stable points at $\mathbf{l} = (\pm 1, 0, 0)$ move to

$$l_y = 0 \ , \quad l_z = \cos i = -4R\Omega_t/[3V(\eta_z - \eta_x)] \ ; \tag{14}$$

these correspond to the stable anomalous retrograde orbits found by Heisler *et al.* (1982) and by Tohline & Durisen (1982). When the figure of the galaxy tumbles somewhat faster, so that $4R/3V\eta_z < \Omega_t < 4R/[3V(\eta_z - \eta_x)]$, the direct pole $\mathbf{l} = (0, 0, 1)$ is still stable, but the two unstable fixed points have merged with the retrograde pole, which is now itself unstable. The two stable points given by equation (14) persist until $\Omega_t$ increases to the point that they merge with the retrograde pole $\mathbf{l} = (0, 0, -1)$; at faster tumbling rates, both poles are stable and there are no other fixed points of precession.

Figure 11 shows the inclination of a disk with the same initial parameters as Figure 9, in a potential with the same parameters but which tumbles at a rate $\Omega_t = 0.07$ about the $z$-axis. All the rings making up the disk precess about that axis, and the the disk settles into the $x$-$y$ plane in a very similar manner to that in the stationary potential.

Since the inclination $i$ of the stably precessing tilted orbits at a given tumbling rate depends on radius according to equation (14), an inclined disk of material which precesses rigidly along with the tumbling potential must have azimuth $\omega = \pm 90°$ and warp according to $\cos i \propto R$. Van Albada, Kotanyi & Schwarzschild (1982) have suggested that the warped and twisted minor-axis dust lanes observed in a number of elliptical galaxies, including NGC 5128 (Centaurus A), can be explained as material which has settled onto this sequence of tilted orbits. Figure 12 shows the development of an initially coplanar disk with $i = 50°$ and azimuth $\omega = 90°$, in the potential of Figure 11, but with equal and opposite tumbling rate; the rings making up this disk indeed settle to the warped shape which precesses along with the potential. The large downward spike in $V_R$ marks the point where the line of nodes of the disk has twisted by a complete revolution as it settled. The rate of settling into this stable warped state is approximately the same as for the disk settling to the equatorial plane.

## 6. Discussion

We have used a set of equations developed by Pringle (1992) to follow the evolution of a viscous twisted disk in a galaxy-like potential which is stationary or tumbling relative to inertial space. These equations represent the result of averaging viscous torques and the gravitational force from the underlying galaxy over the orbit of each gas element in the disk. In an axisymmetric potential, we find that the disk settles to the equatorial plane at a rate determined largely by the coefficient $\nu_2$ associated with shear perpendicular to the local disk plane. If the disk is initially close to the galaxy equator, the settling depends on viscosity and the rate of differential precession in a way which is well described by an analytic formula given by Steiman-Cameron & Durisen (1990). In a highly inclined disk, however, 'breaking waves' of curvature steepen as they propagate through the disk, up to the maximum gradient allowed by the radial grid; angular momentum and mass are lost from the calculation where the advective speed $V_{adv}$ changes sign, and the computed curvature does not approach a smooth limit as the grid is made finer.

The predictions of Pringle's (1992) orbit-averaged equations are somewhat at variance with those obtained in simulations using smoothed-particle hydrodynamics. According to the orbit-averaged equations, disks settling in oblate and prolate potentials are equivalent apart from a sign change in the azimuthal angle; in particular, stably precessing states which have no sense of spirality will be identical in oblate and prolate potentials. The different behavior of viscous disks in oblate as opposed to prolate potentials found by Christodoulou *et al.* (1992) and Katz & Rix (1992) must depend on aspects of their dissipative scheme which are not reflected in the orbit-averaged equations. This is surprising since the particles in those calculations follow nearly circular orbits, so that one would expect orbit-averaging to give a good representation of the dynamics. In particular, we find that viscosity cannot lock a tilted ring into a steadily precessing warped state with inclination $i$ given by $\cos i \propto R$. This configuration is neutrally stable for a freely precessing disk, but according to the orbit-averaged scheme, it is destabilized by viscosity.

In a triaxial potential which is stationary in inertial space, stable orbits exist about both the longest and shortest axes, but not about the intermediate axis; an orbit which is not in one of the symmetry planes must precess about either the long or the short axis. We find that, as expected, an inclined disk settles into a plane perpendicular to that about which the individual gas orbits precess. The inclination at each radius of a precessing disk must vary with azimuth as the ring precesses; this could potentially be used as a diagnostic for triaxiality in the potentials of galaxies such as NGC 4753 (Steiman-Cameron *et al.* 1992) in which a tilted gas disk has been twisted by precession. We find that the disk settles somewhat faster than in an axisymmetric galaxy, since it twists simultaneously about two perpendicular axes.

If the figure of the potential tumbles about one of its principal axes, there is a stable tilted orbit at each radius (up to some maxiumum), precessing so as to remain stationary relative to the underlying potential. In a system tumbling about the shortest axis, these orbits have angular momentum which is retrograde with respect to the figure tumbling; while an initially prograde viscous disk in such a potential settles into the plane normal to the short axis, an initially retrograde disk can settle towards the warped disk, composed of orbits which at each radius precess so as to remain stationary relative to the underlying galaxy. The time required to settle into such a warped state is approximately the same as that for settling to a principal plane of the potential.

## Acknowledgements

We are grateful to Jeremy Goodman for his useful comments. WNC would like to thank the Fannie and John Hertz Foundation for their gracious and continued support through a Fellowship. LSS was partially supported by NSF grants AST 90-20650 and AST 93-20403, and thanks the Institute for Advanced Study for hospitality and support through an AMIAS fellowship during the period when this work was begun.

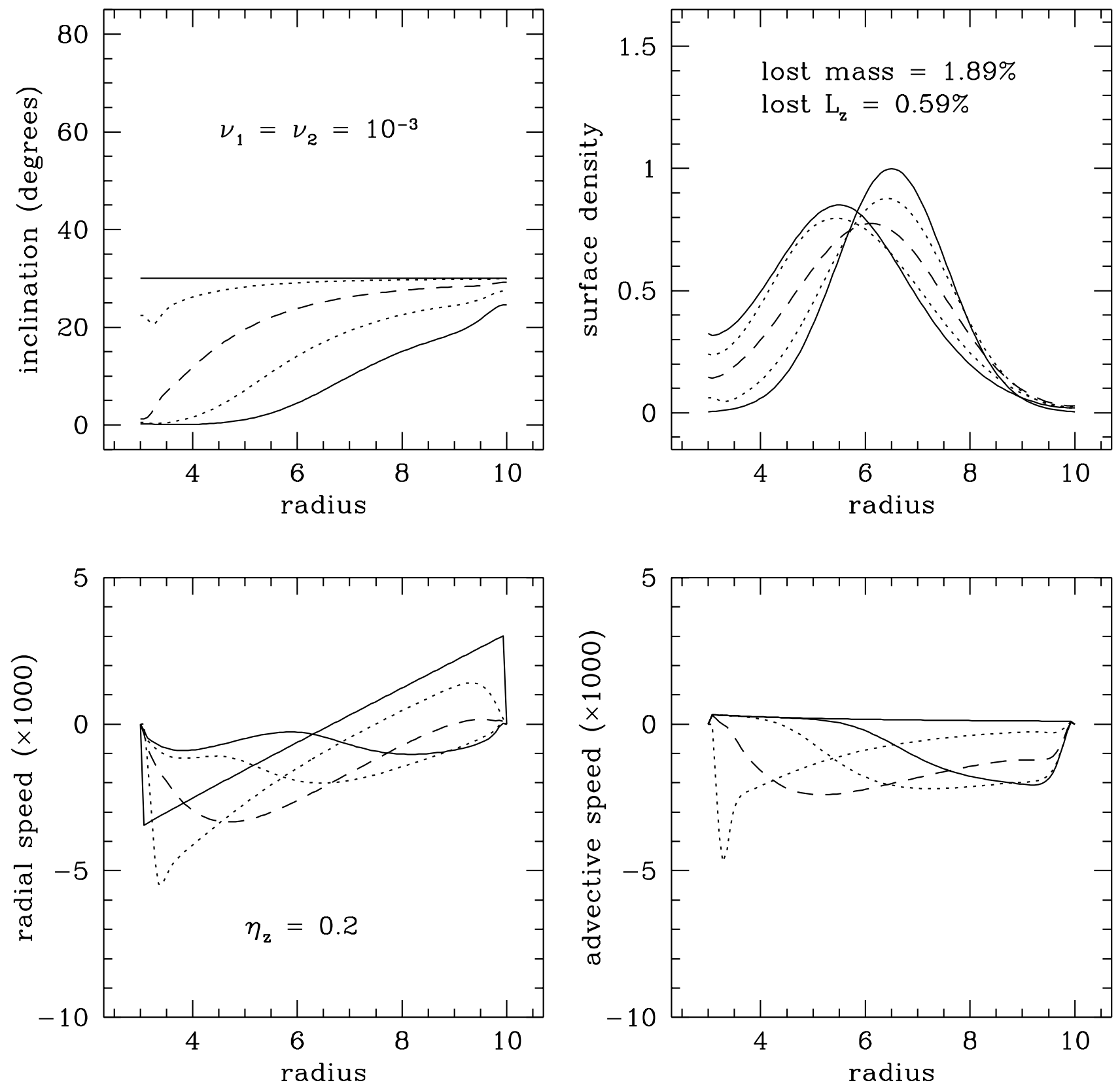

Fig. 1.— Settling of an initially coplanar disk inclined at 30° to the equator of an oblate galaxy potential corresponding to $V = 250\, km\ s^{-1}$, $\eta = 0.2$: the viscous coefficients $\nu_1 = \nu_2 = 10^{-3}$ and the computational grid contains 100 rings equally spaced in radius. The top left panel shows the inclination, which at each radius decreases with time; the top right shows the density, the lower left $V_R$, and the lower right $V_{adv}$. The different line types correspond to equally spaced time intervals. The time $t = 600$ at the end of the run corresponds to 6Gyr, and the timestep was 0.01, corresponding to $10^5$ years.

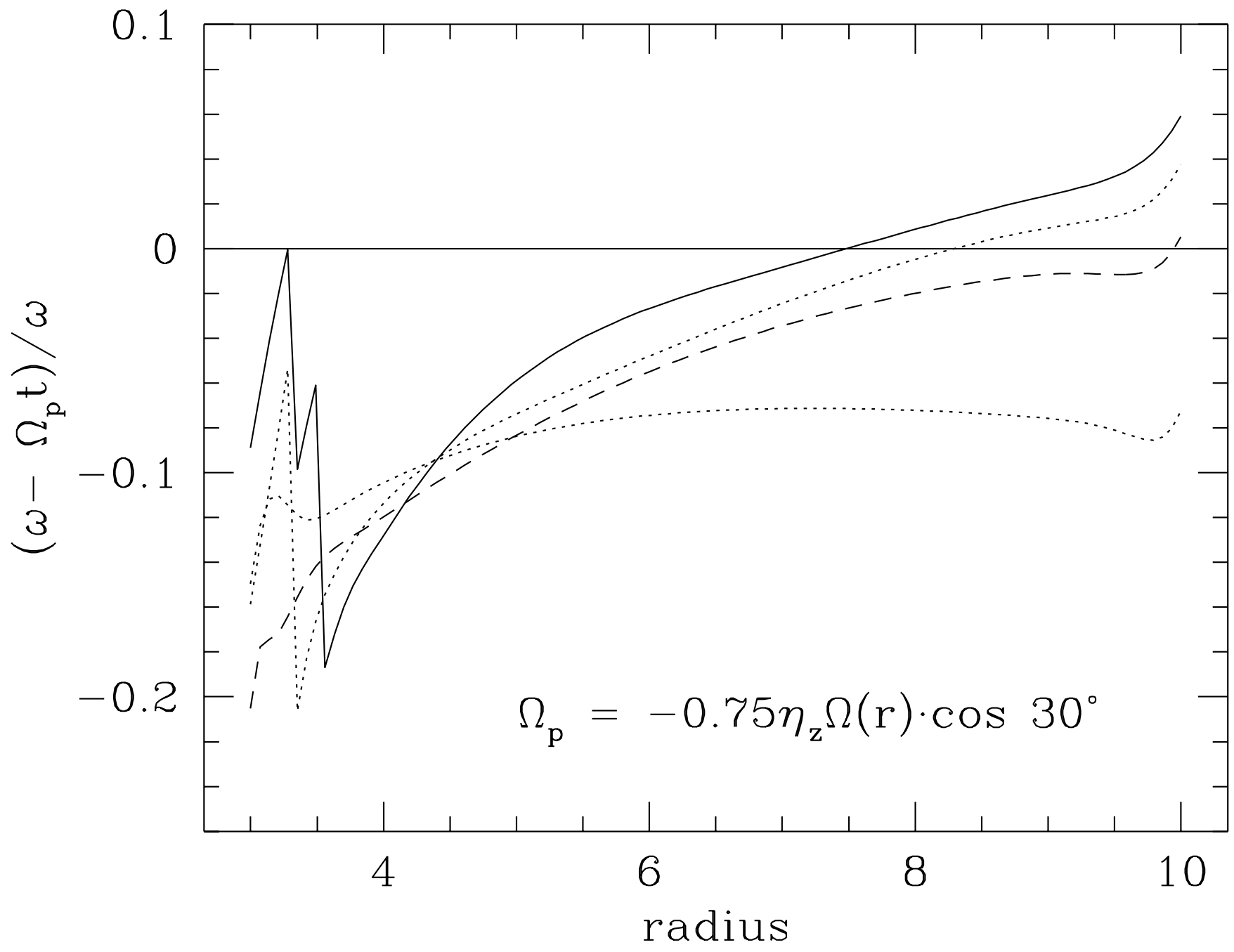

Fig. 2.— For the computation shown in Fig 1, the difference between the azimuth $\omega$ of the line of nodes of the disk at radius $R$, and that expected in free precession at an inclination of 30°. The different line types correspond to the same times as in Fig 1.

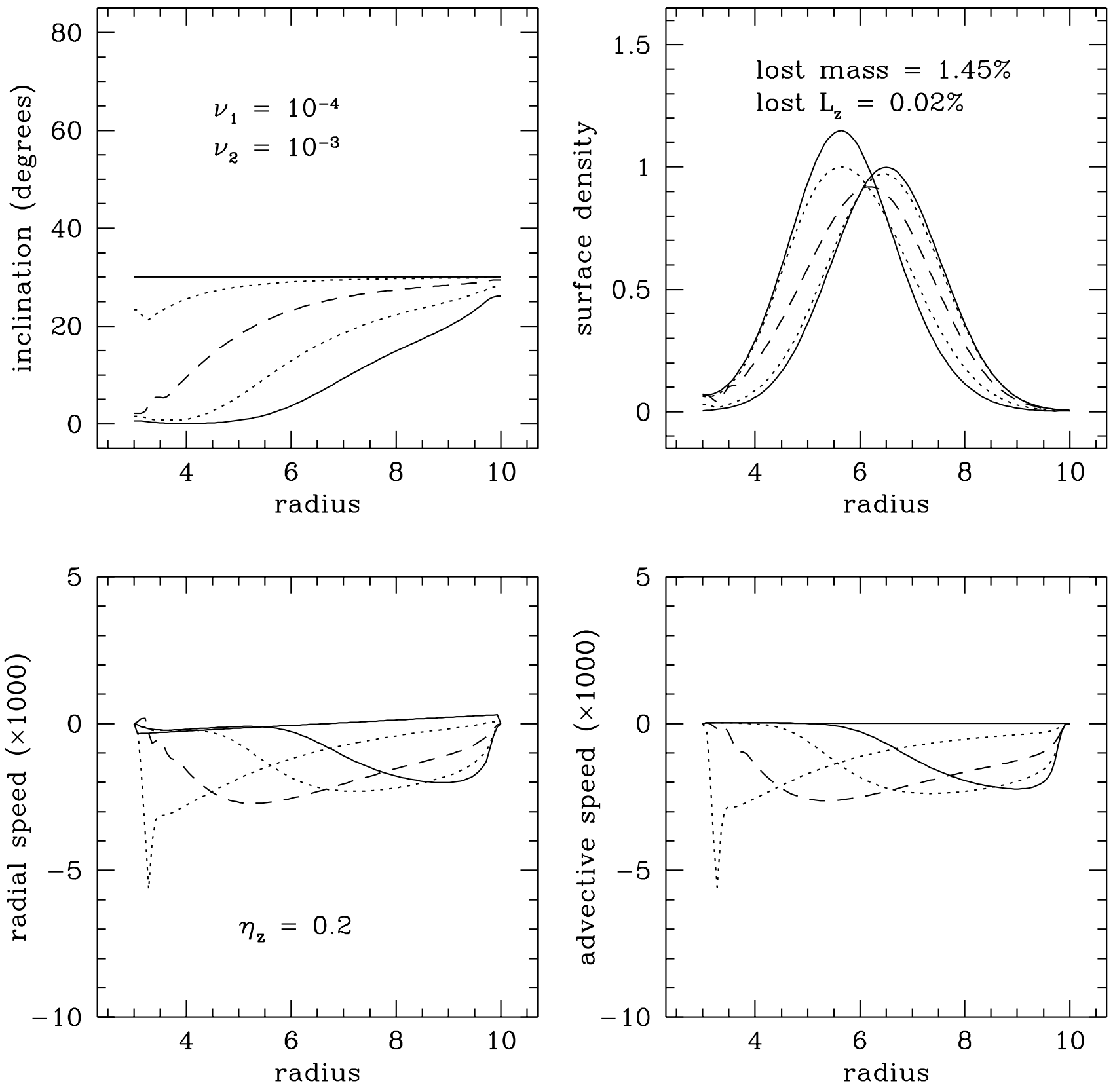

Fig. 3.— As Fig. 1, but for $\nu_1 = 10^{-4}, \nu_2 = 10^{-3}$.

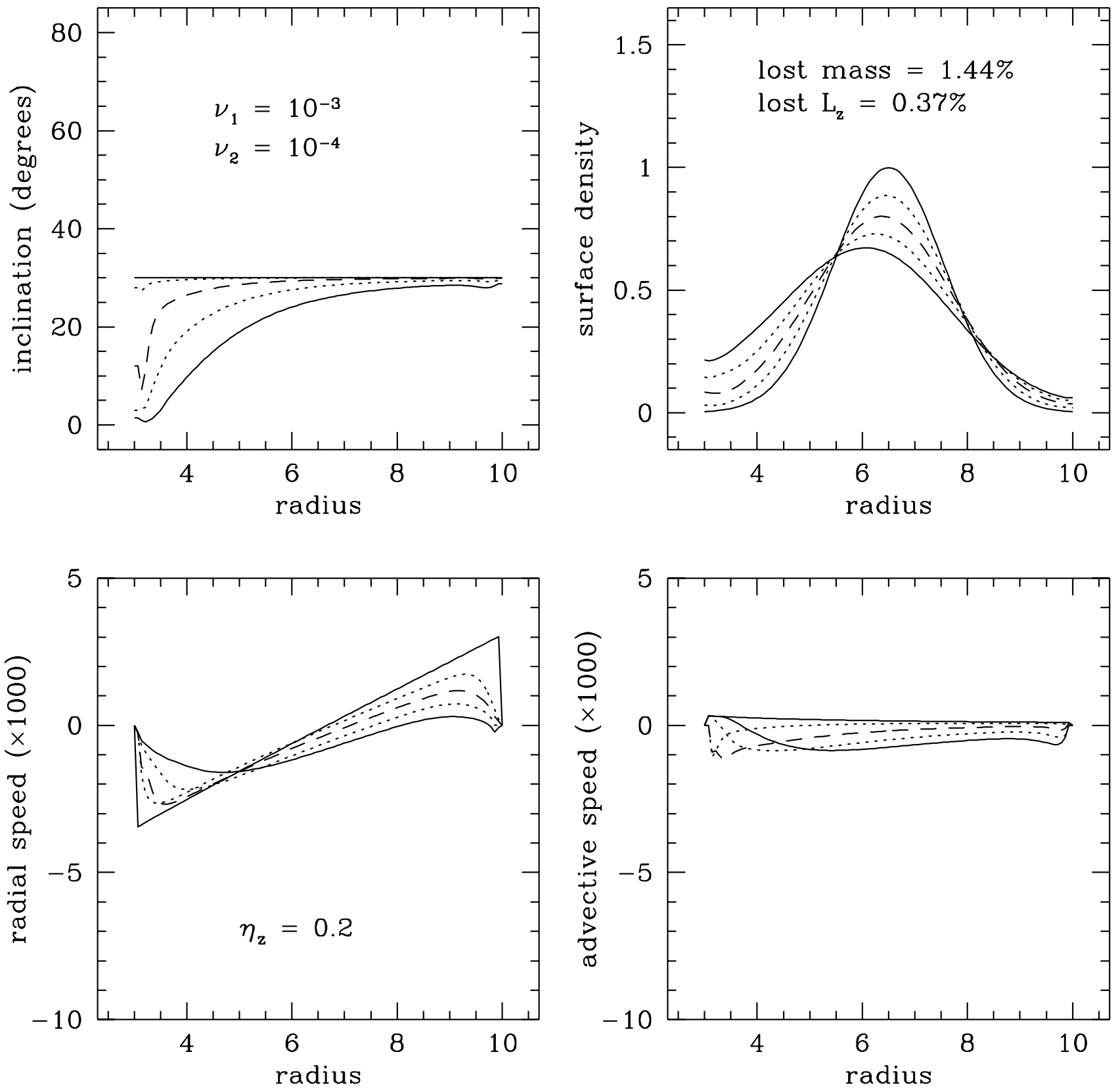

Fig. 4.— As Fig. 1, but for $\nu_1 = 10^{-3}, \nu_2 = 10^{-4}$.

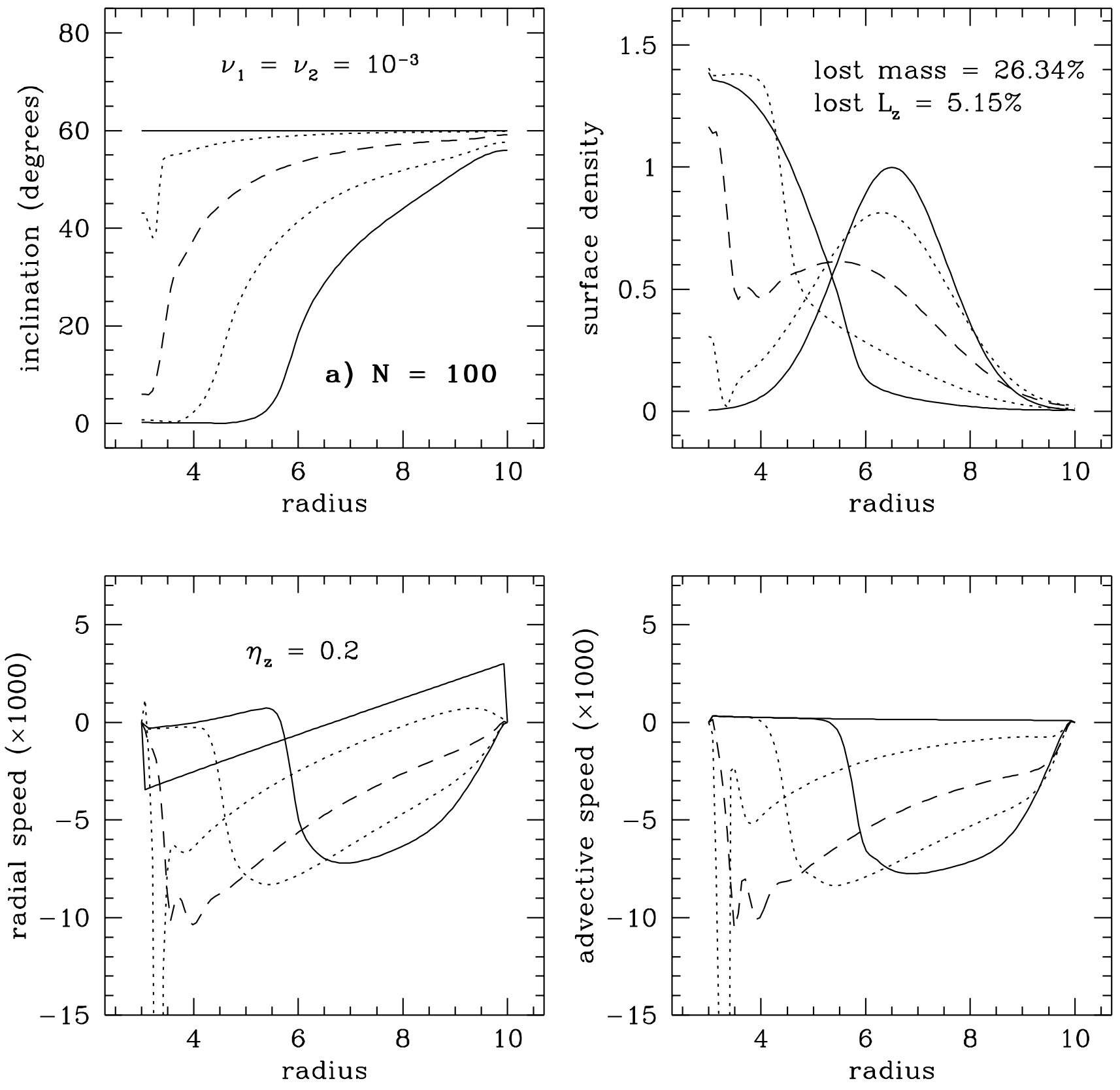

Fig. 5.— a) as Fig. 1, but for an initial inclination of 60°, and a run yr time 50% longer, corresponding to 9Gyr.

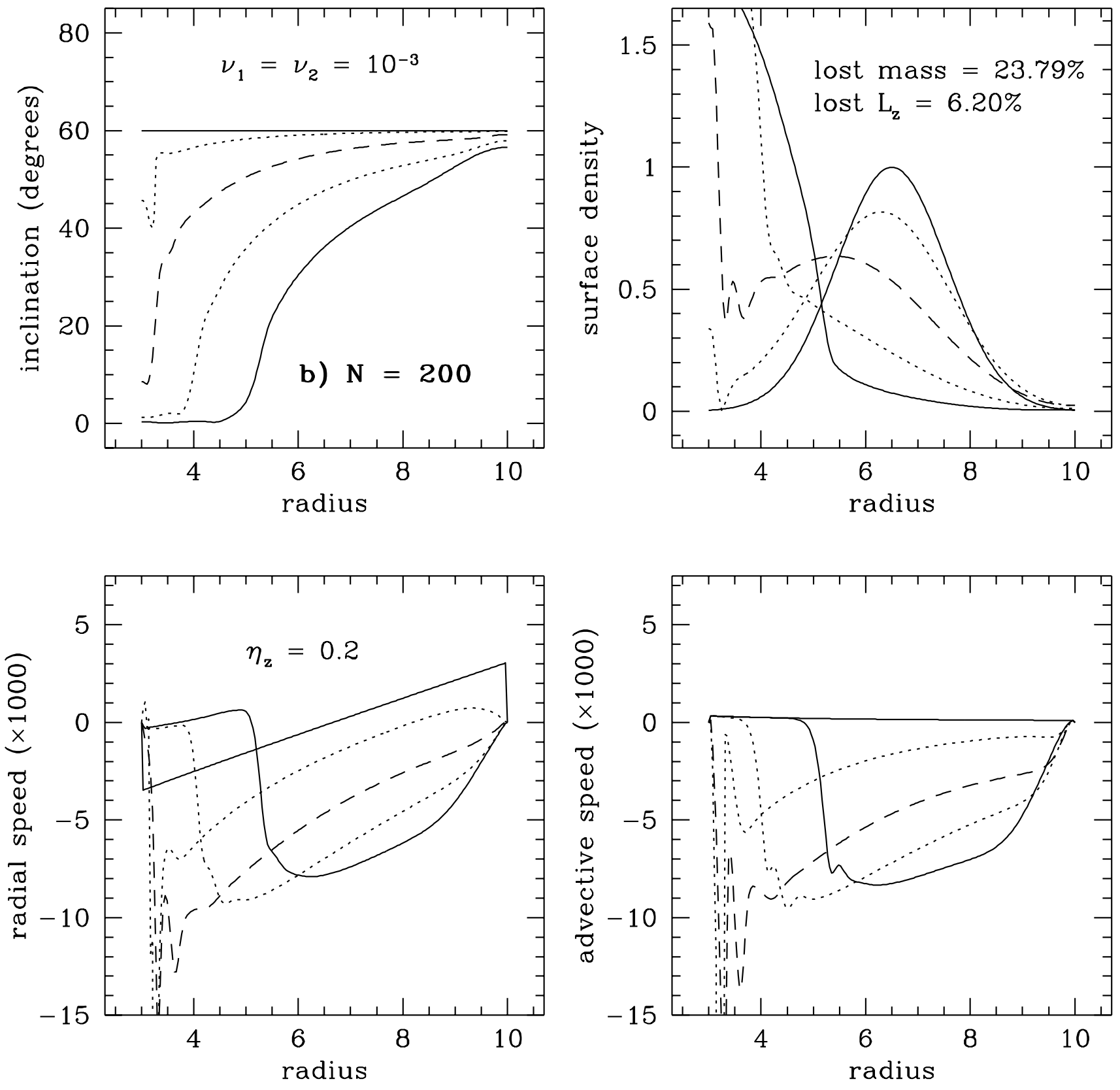

Fig. 5.— b) The same computation run as in 5a), but with a finer grid with 200 equally spaced rings.

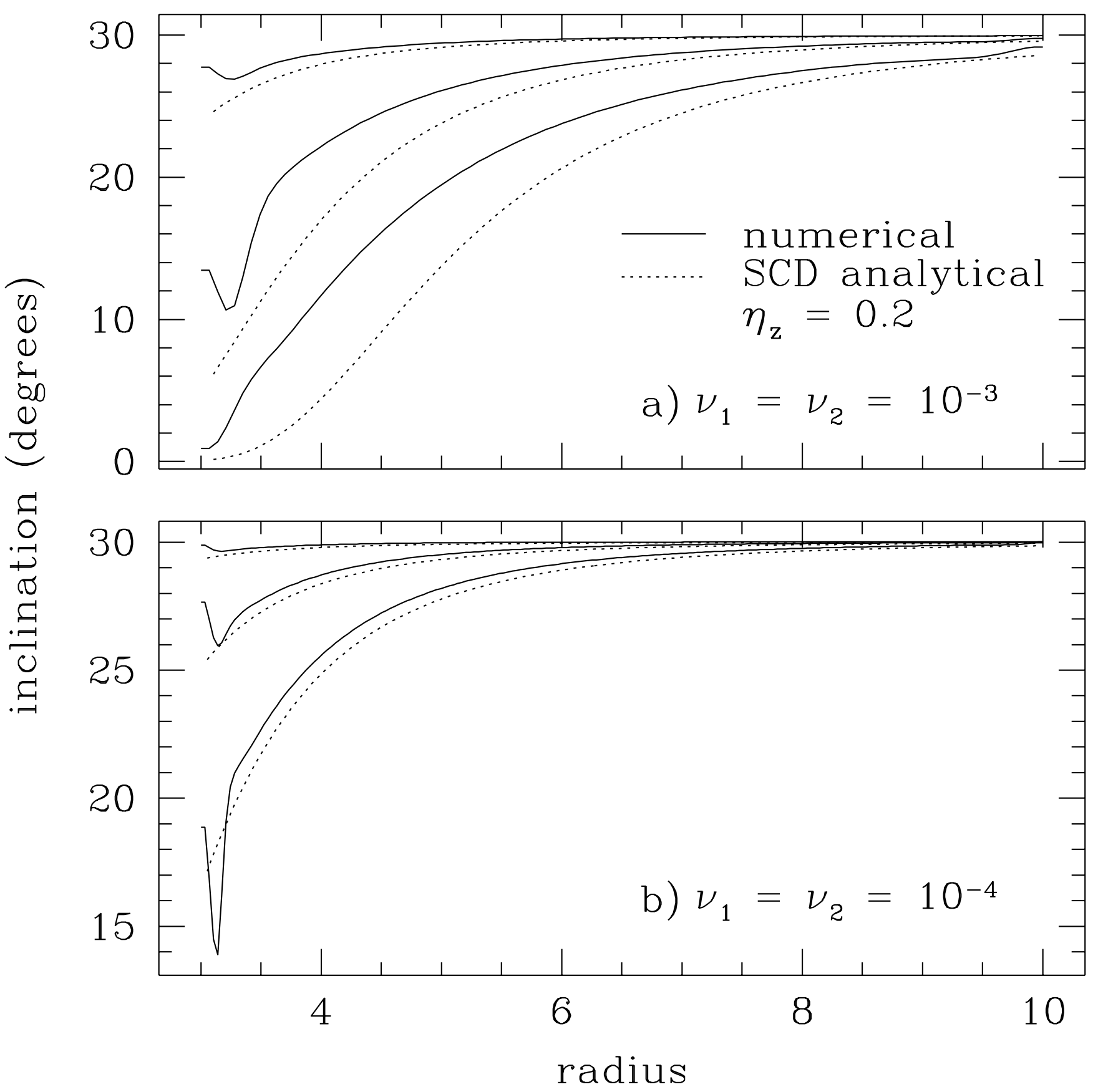

Fig. 6.— The run of inclination with radius from our computations (solid lines), compared with the prediction of equation (8) given by the dashed lines. Panel a) is for the parameters of Fig. 1, and b) is for viscosity parameters ten times smaller.

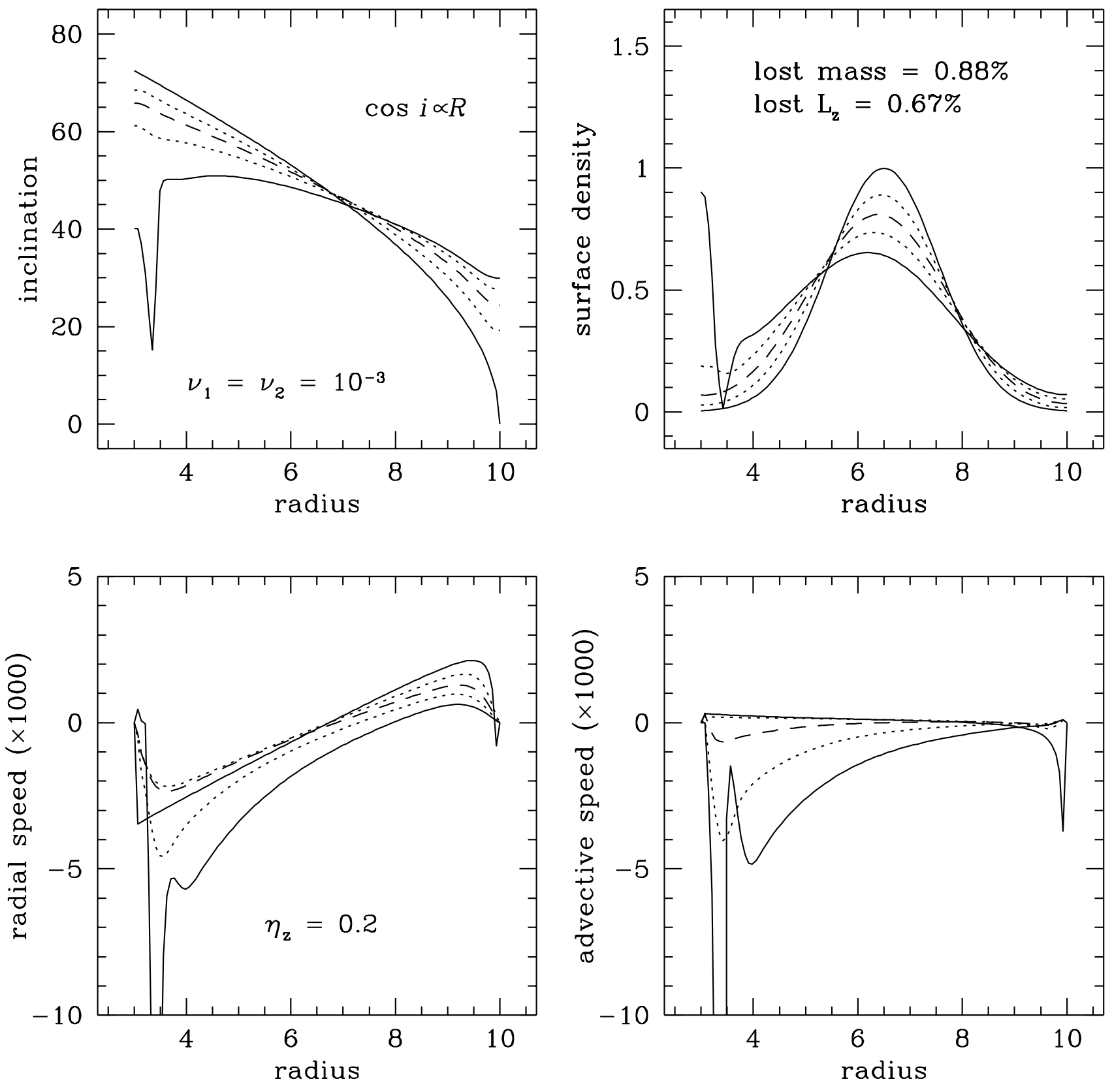

Fig. 7.— As Fig. 1, but with a disk which initially is warped following $\cos i \propto R$, such that it would precess freely at the rate $\Omega_t = -0.035$, or $3.5\,km\ s^{-1}$ per kiloparsec, in the absence of viscosity. This configuration is destabilized by viscosity.

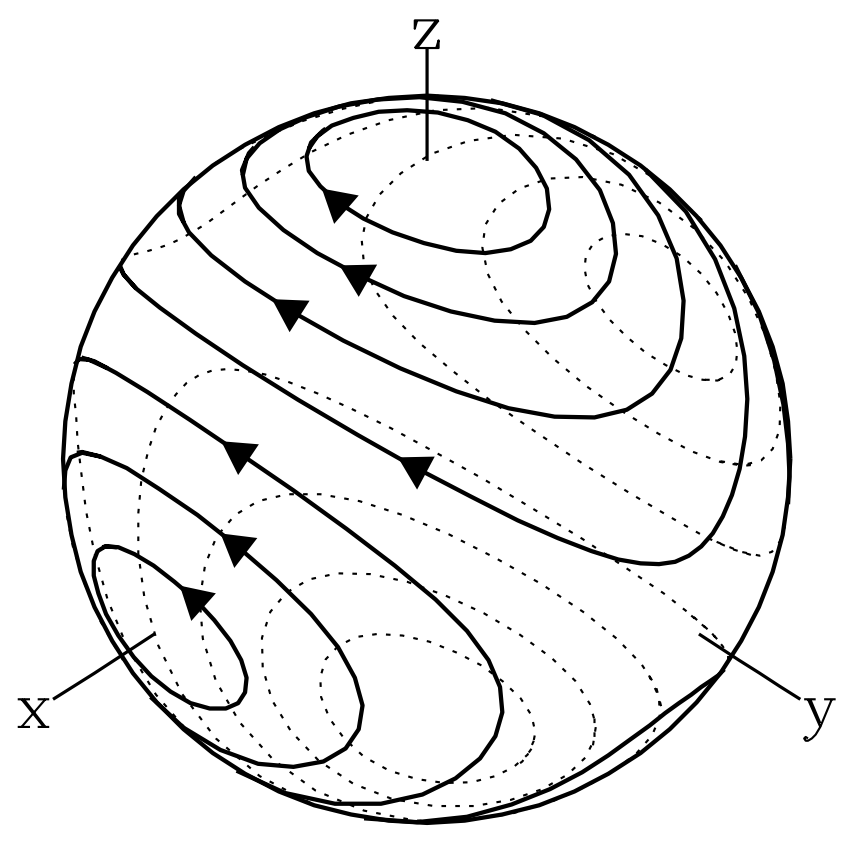

Fig. 8.— Free precession trajectories in the triaxial galaxy potential of equation (10); the path of the ring normal **l** is plotted on the unit sphere.

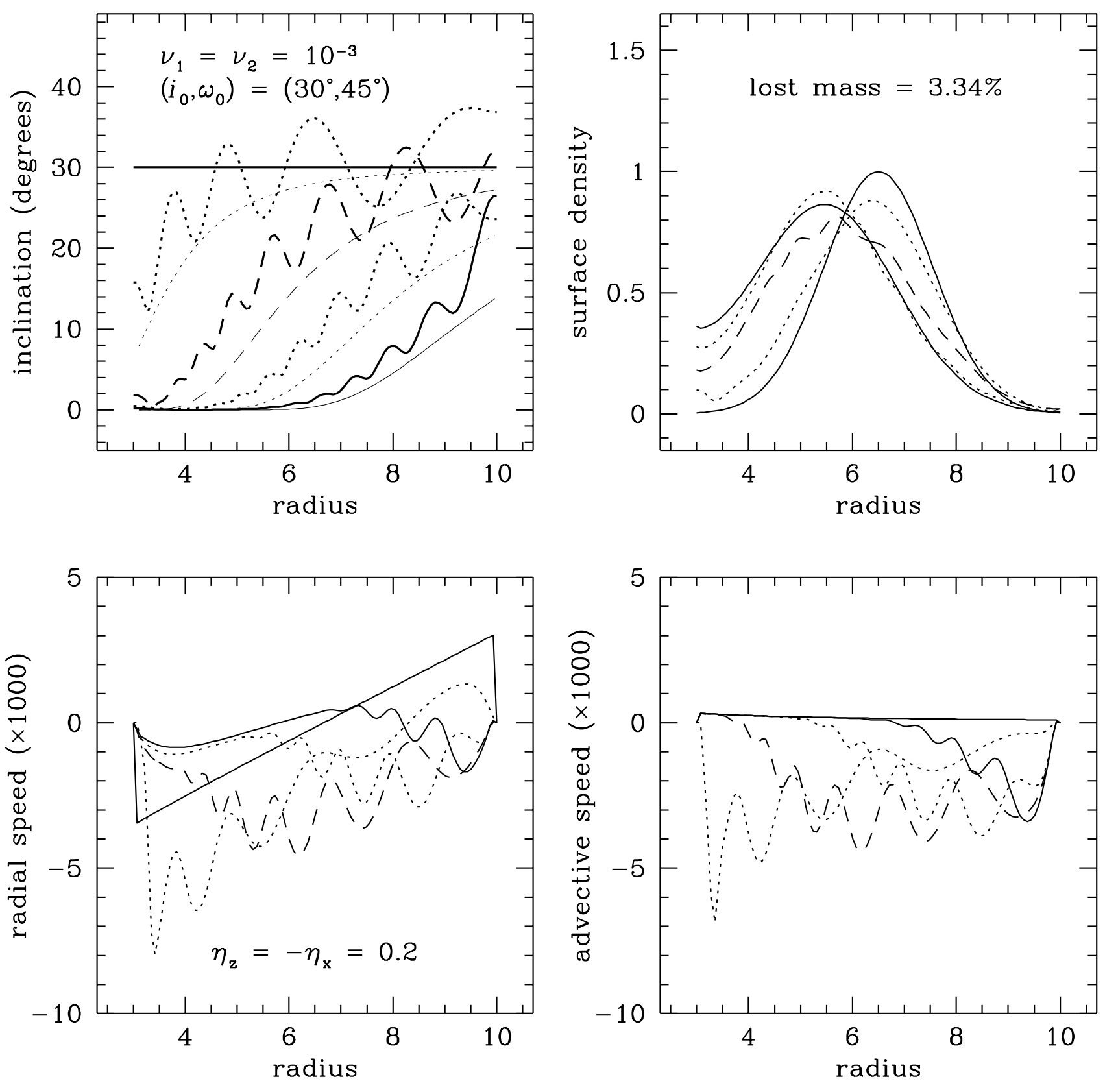

Fig. 9.— Settling of an initially planar disk in a triaxial potential with $\eta_x = -0.2$, $\eta_z = 0.2$; the disk starts at $i = 30°$, $\omega = 45°$, and other parameters are as in Fig. 1. The prediction of equation (8 ), as discussed in the text, is shown at each time by the lighter, monotonic, curve of the same line type.

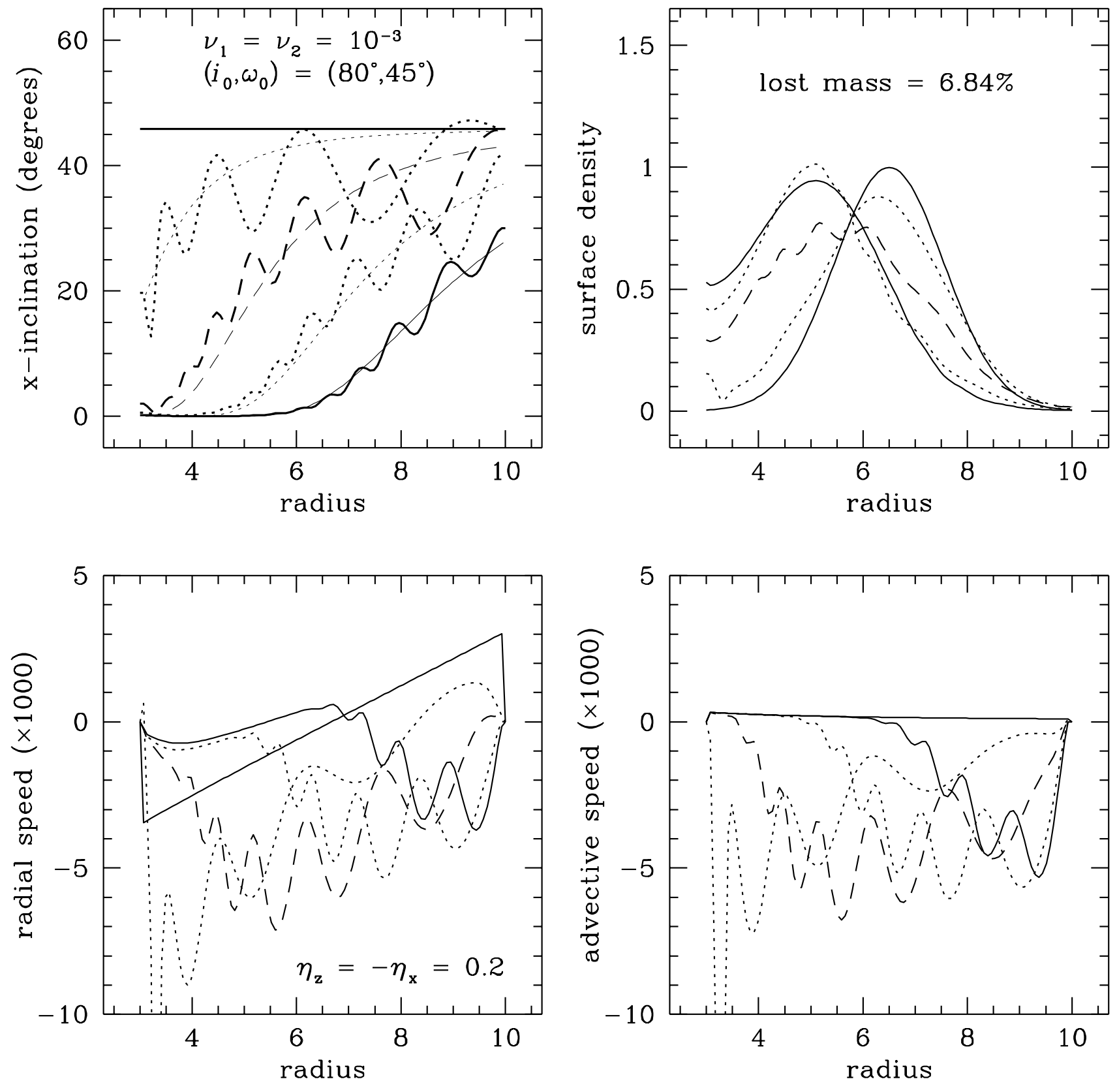

Fig. 10.— As Fig. 9, except that the inclination to the $x$-axis is shown, for a disk starting with $i = 80°$, $\omega = 45°$. The disk precesses about the $x$-axis and eventually settles into the $y$-$z$ plane, approximately following the prediction of equation (8 ).

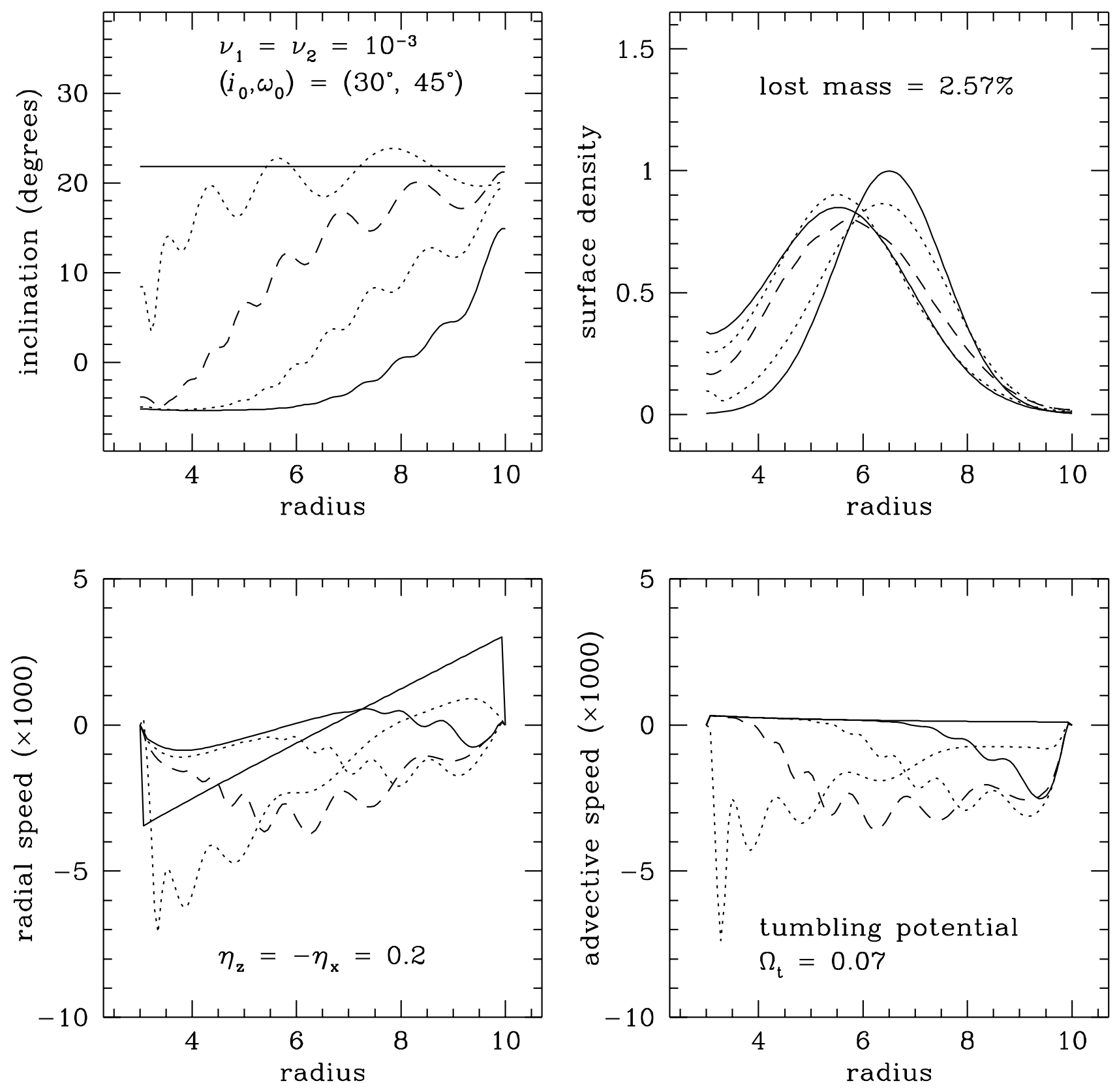

Fig. 11.— As Fig. 9, but the galaxy potential now tumbles about the $z$-axis at a rate $\Omega_t = 0.07$, corresponding to $7\,km\ s^{-1}$ per kiloparsec.

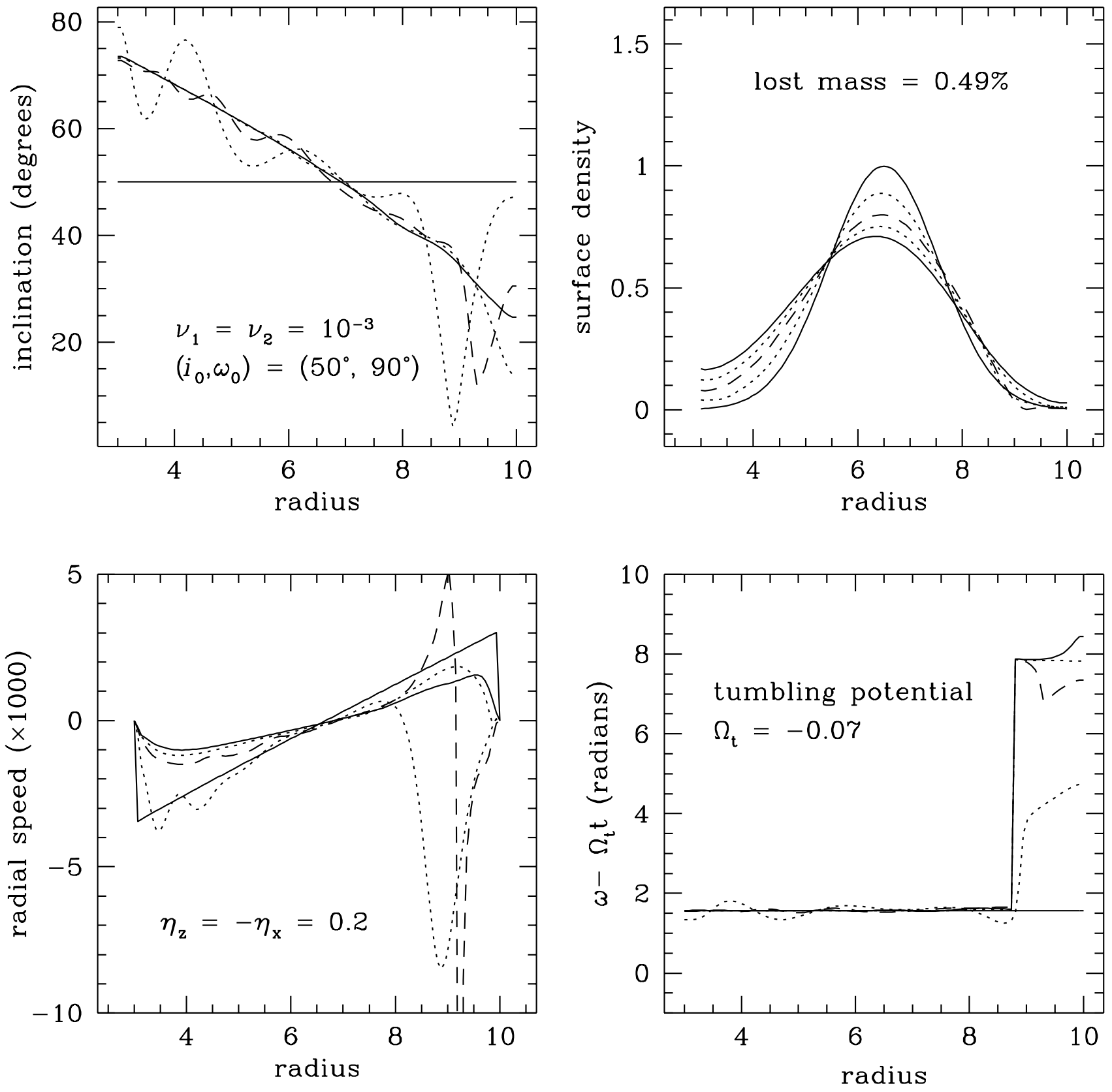

Fig. 12.— The inclination to the $z$-axis, surface density, inflow speed $V_R$ and azimuth of the line of nodes relative to the longest axis of the tumbling potential, for a disk initially inclined at 50° to the $x$-$y$ plane and rotating retrograde with respect to the tumbling of the galaxy, which is given by $\Omega_t = -0.07$. At the start the relative azimuth is $\omega = 90°$.